\documentclass{ws-procs9x6}
\newcommand{\mrm}[1]{\mbox{\rm #1}}
\newcommand{\nn}{\nonumber}
\newcommand{\be}{\begin{equation}}
\newcommand{\bea}{\begin{eqnarray}}
\newcommand{\eea}{\end{eqnarray}}
\newcommand{\ee}{\end{equation}}

\newcommand{\D}{D\hspace{-8pt}\slash}
\newcommand{\dv}{\partial\hspace{-7pt}\slash}

\def\cal{\fam2 }

\begin{document}

\title{Testing the Leptogenesis Mechanism of the Seesaw Model}

\author{ELIZABETH JENKINS}

\address{ Department of Physics, \\
University of California at San Diego, \\ 
9500 Gilman Drive, \\
La Jolla, CA 92093, USA \\ 
E-mail: ejenkins@ucsd.edu}

\maketitle

\abstracts{
The seesaw theory, the leading theory for particle interactions, provides a viable mechanism for generating the matter-antimatter asymmetry of the universe.
Testing the leptogenesis mechanism directly
requires measurement of the $d=6$ operator of the low-energy effective Lagrangian, in addition
to the more familiar $d=5$ operator which generates Majorana masses for the light neutrinos when the electroweak symmetry is spontaneously broken.  This important experimental challenge awaits the next generation of particle physicists.}

We are at a very interesting crossroads in particle physics.  Experiment has now established that neutrinos have mass, and that the neutrino mass eigenstates, $\nu_1$, $\nu_2$ and $\nu_3$ with 
masses $m_1 < m_2 < m_3$,
are not the same as the neutrino flavor eigenstates $\nu_e$, $\nu_\mu$ and $\nu_\tau$ which appear in weak doublets with the left-handed charged leptons $e^-$, $\mu^-$ and $\tau^-$. 
We know quite a bit about the parameters of neutrino mixing from atmospheric and solar neutrino oscillations.  What I wish to focus on here is what we do not know about the parameters of the lepton sector.

Our leading candidate theory at high energies is the seesaw model,\cite{seesaw} originally proposed 24 years ago.  The seesaw model is the minimal extension of the Standard Model (SM) which adds right-handed neutrinos to its particle content.
Including all renormalizable interactions which are allowed by $SU(3) \times SU(2)_L \times U(1)_Y$ gauge symmetry yields the seesaw Lagrangian.
Since right-handed neutrinos are gauge singlets, Majorana mass terms for these neutrinos are allowed,
and unlike other fermions, they have a mass $M$ which is much larger than the weak scale and which is independent of electroweak symmetry breaking (EWSB). 
The only interactions of the right-handed neutrinos are Yukawa interactions with the Higgs scalar doublet and the lepton doublets.  Thus, the leptonic Lagrangian of the high-energy seesaw theory is given by
\begin{eqnarray}
{\cal L}_{\rm seesaw} &=& i \, \overline{\ell_L} \,
\D \, \ell_L + i \, \overline{e_R} \, \D \, e_R
+ i \, \overline{N_R} \, \dv \, N_R \nn\\
&&-  \overline{\ell_L} \,{\phi} \, {Y_e}  \, e_R
- \overline{\ell_L} \,{\widetilde\phi} \, {Y_\nu}  \, N_R
-\frac{1}{2}\,\overline{{N_R}^c} \, M \,{N_R}
+ \mrm{h.c.} 
\end{eqnarray}
where $\ell_{L\alpha}$ are lepton electroweak doublets, $e_{R\alpha}$ are charged lepton singlets,
$\phi$ is the Higgs doublet and $N_R$ are heavy neutrino singlets.  For further explanation of notation, see Ref.~\refcite{bgj}.

Let me briefly remind you of the attractive features of the seesaw model.  First and foremost, the lightness of weakly-interacting neutrinos and the heaviness of sterile neutrinos is explained by
the seesaw mechanism.  The weakly-interacting neutrinos acquire small Majorana masses after the electroweak symmetry breaks.  These Majorana masses are $SU(2)_L$ triplet and require two insertions of the Higgs vacuum expectation value $v/\sqrt{2}$.  Thus, the light neutrino Majorana masses are order $(v^2/2M)$ and are
suppressed by a factor of $(v/\sqrt{2}M)$ relative to the Dirac masses, proportional to $v/\sqrt{2}$, of the other SM fermions.  Hence, the large hierarchy between the light neutrino masses and the rest of the SM fermions
can be understood.

A second attractive feature of the seesaw model is that
leptogenesis occurs at high energy.\cite{fy}  The heavy right-handed neutrinos are unstable to decay to lepton and scalar (Higgs) doublets.  Interference between tree and loop decay diagrams of the right-handed neutrinos leads to $CP$ violation in the decay processes. 
Thus, a small lepton asymmetry is generated at high energies in heavy neutrino decay.  This leptogenesis is characterized by the $CP$-violating asymmetry of the lightest right-handed neutrino $N_1$ with mass $M_1$, and can be written in terms of the neutrino Yukawa matrix combination $Y_\nu^\dagger Y_\nu$ and the mass ratios of the heavy Majorana neutrinos
$M_2/M_1$ and $M_3/M_2$. 

Other attractive aspects of the seesaw model are that it is a renormalizable high-energy theory, which implies that
it has a finite number of high-energy parameters; it is natural in the sense
of `t Hooft,\cite{tHooft}  i.e. all renormalizable interactions allowed by the SM gauge symmetry are included in the Lagrangian; experiment indicates a seesaw scale $M$ consistent with gauge coupling unification; and it is easy to embed the seesaw model in a grand unified or partially unified
theory.  With such an embedding, asymmetries of the neutrino field content and of the gauge symmetry 
of the SM are low-energy artifacts of spontaneous symmetry breakdown of a more symmetric theory.      

Although the seesaw model is not the only possible theory of neutrino mass, it is by far the theoretically favored one.  A logical alternative to the seesaw model is to add right-handed neutrinos, but to suppose that neutrinos obtain only Dirac masses from Yukawa couplings upon EWSB, just like all the other SM fermions.  This
supposition requires that lepton number $L$ is a globally conserved quantum number so that
Majorana mass terms are forbidden.  Such an assumption is unnatural in the technical sense.
In this case, in contrast to the seesaw theory, there is no understanding of why light
neutrino masses are so much smaller that the masses of all the other SM fermions.  Furthermore, there is no leptogenesis at high energy if $L$ is exact, so the matter-antimatter asymmetry of the universe
is not addressed.  These are strong theoretical reasons for favoring the seesaw model.  To summarize:
theorists believe that neutrinos are Majorana particles.

In the seesaw model, the
lepton asymmetry generated at high energies is thought to be reprocessed by non-perturbative processes at the electroweak scale, resulting in a baryon-antibaryon asymmetry.  As is well-known, it is possible to generate a sufficient lepton asymmetry at high energy for the leptogenesis $\rightarrow$ baryogenesis scenario to provide the solution of the matter-antimatter asymmetry of the universe.\cite{lepto}  It is worth pointing out that the magnitude (and sign!) of the baryon-antibaryon asymmetry of the universe is known quite precisely from the WMAP data, so the leptogenesis constraint
is significant.

This state of affairs is quite interesting.  We have a viable and compelling theory of leptogenesis.  
It needs to be tested.  Can we establish experimentally that leptogenesis $\it is$ the solution of the matter-antimatter asymmetry puzzle?  The characteristic energy scale of the seesaw theory is such that we may never build an accelerator which can directly probe its physics.  Can we establish the seesaw theory by performing measurements at accessible energies?

At ``low'' energies, the heavy right-handed neutrinos are not in evidence, so we can measure only the parameters of the low-energy effective theory in which the heavy Majorana neutrinos are integrated out of the theory.  The effective Lagrangian has the form
\begin{equation}
{\cal L}_{\rm eff} = {\cal L}_{SM} + \delta {\cal L}_{d=5} + \delta {\cal L}_{d=6} + \cdots ,
\end{equation}
where higher dimensional Lagrangian terms with ${d > 4}$ are suppressed by the power $1/ M^{d-4}$ of the seesaw scale.  The $d=5$ and $d=6$ operators of the low-energy effective seesaw theory are\cite{bgj}
\begin{eqnarray}
\delta {\cal L}_{d=5} &=& \frac{1}{2}\, c_{\alpha \beta}^{d=5} \,
\left( \overline{{\ell_{L\alpha}}^c} \tilde \phi^* \right) \left(
\tilde \phi^\dagger \, \ell_{L \beta} \right) + {\rm h.c.}, \\
\delta {\cal L}_{d=6} &=& c^{d=6}_{\alpha \beta} \, \left( \overline{\ell_{L\alpha}} \tilde \phi
\right) i \dv \left( \tilde \phi^\dagger \ell_{L \beta} \right),
\end{eqnarray}
where the coefficients $c^{d=5}$ and $c^{d=6}$ are related to the high-energy parameters of the seesaw theory $Y_\nu$ and $M$ by
\begin{eqnarray}
c^{d=5} &=& Y_\nu^* \, (M^*)^{-1} \, Y_\nu^\dagger \ , \\
c^{d=6} &=& Y_\nu \, (|M|^2)^{-1} \, Y_\nu^\dagger. 
\end{eqnarray}
Measurement of both the $d=5$ and $d=6$ operator coefficients suffices for construction of the high-energy seesaw Lagrangian from the low-energy data, when the number of right-handed neutrinos is equal to the number of light fermion generations,\cite{bgj} which is the most attractive situation theoretically.  (The formulae making the connection between the low-energy and high-energy parameters are involved\cite{bgj} and are suppressed here.  The important point is that this connection is possible.)

For three generations of fermions, there are 15 real parameters and 6 phases in the high-energy seesaw Lagrangian.  Using
chiral symmetries, it is possible to define these physical parameters in a standard high-energy 
basis\cite{bgj}  
\begin{eqnarray}
Y_e &=&{{\sqrt{2}} \over v} {\rm diag}(m_e, m_\mu, m_\tau), \qquad
M = {\rm diag}(M_1, M_2, M_3), \\
Y_\nu &=& \left(\begin{array}{ccc}
u_{e1} e^{i \Phi_{e1}} & u_{e2} e^{i\Phi_{e2}} & u_{e3} e^{-i(\Phi_{e1} + \Phi_{e2})} \\
u_{\mu1} e^{i \Phi_{\mu1}} & u_{\mu2} e^{i\Phi_{\mu2}} & u_{\mu3} e^{-i(\Phi_{\mu1} + \Phi_{\mu2})} \\
u_{\tau1} e^{i \Phi_{\tau1}} & u_{\tau2} e^{i\Phi_{\tau2}} & u_{\tau3} e^{-i(\Phi_{\tau1} + \Phi_{\tau2})} \\
\end{array}\right),
\end{eqnarray}
with all phases appearing in the Yukawa matrix $Y_\nu$.  The low-energy effective theory including the $d=5$ and $d=6$ operators also contains 15 real parameters and 6 phases.  Three of the real parameters are the charged lepton masses, and are known. 
The $d=5$  and $d=6$ operator coefficients each contain 
6 real parameters and 3 phases.  
  
The $d=5$ operator is the usual Weinberg operator.\cite{weinberg}  Upon electroweak symmetry breaking (EWSB), it generates a Majorana mass term for the weakly-interacting light neutrinos with mass matrix
\begin{equation}
m_{\alpha\beta} \equiv  -\frac{1}{2} v^2 c^{d=5}_{\alpha \beta}
\end{equation}
in the flavor basis.  Diagonalization of this matrix yields the light neutrino masses $m_1$, $m_2$, $m_3$ and the lepton mixing matrix\cite{PMNS}
\begin{equation}
U_{PMNS} = U_{CKM}\ 
\left(\begin{array}{ccc}
e^{i\phi_{1}/2}& 0 & 0\\
0& e^{i\phi_{2}/2}& 0\\
0& 0 & 1  \\
\end{array}\right),
\end{equation}
\begin{equation}
U_{CKM}=
\left(\begin{array}{ccc}
c_{12} c_{13} & s_{12} c_{13} & s_{13}\, e^{-i\delta} \\
-s_{12} c_{23} - c_{12} s_{23} s_{13} \, e^{i\delta}
&c_{12} c_{23} - s_{12} s_{23} s_{13} \, e^{i \delta}
&s_{23} c_{13} \\
s_{12} s_{23} - c_{12} c_{23} s_{13} \, e^{i\delta}
&-c_{12} s_{23} - s_{12} c_{23} s_{13} \, e^{i\delta}
&c_{23} c_{13} \\
\end{array}\right) .
\end{equation}
Notice that, in contrast to the quark CKM mixing matrix, the lepton PMNS mixing matrix, contains 3 $CP$-violating phases, since phase redefinitions on the Majorana neutrinos are not allowed.\footnote{I am working in the convention
that the Majorana neutrino fields satisfy $\nu_i^c = \nu_i$.\cite{bgj}}  
Some of these parameters ($\theta_{12}$, $\theta_{23}$, and the square mass differences
$\Delta m_{21}^2 \equiv m_2^2 - m_1^2$ and $\Delta m_{32}^2 \equiv m_3^2 - m_2^2$) already
have been observed in low-energy neutrino oscillations.
The presently undetermined parameters, namely $s_{13}$, $\delta$ and the absolute scale of light neutrino mass $\bar m$, are the focus of upcoming experiments, and undoubtably will be observed 
or severely constrained in the coming decade.

The $d=6$ Lagrangian usually is not considered 
in the low-energy effective seesaw theory because it is suppressed by an additional power of the seesaw scale.  This suppression factor is significant since the seesaw scale is large compared to the weak scale.  Upon EWSB, the $d=6$ operator generates a flavor-nondiagonal $d=4$ contribution to the kinetic energy term of the light weakly-interacting neutrinos given by
the Hermitian matrix 
\begin{equation}
\lambda_{\alpha \beta} \equiv \frac{1}{2} v^2 c^{d=6}_{\alpha \beta} 
\end{equation} 
in the flavor basis.
Normalization of the effective neutrino kinetic energy term and diagonalization of the kinetic  energy and Majorana mass terms implies an effective lepton mixing matrix\cite{bgj}
\begin{equation}
U_{\rm eff} =\left({\mathbb I} - \Omega \frac{\lambda}{2} \Omega^\dagger \right) U \, ,
\end{equation}
which is no longer unitary due to effects of the $d=6$ operator. 
($\Omega$ represents a rephasing of the charged lepton fields which is defined in Ref.~\refcite{bgj}.
Understanding this subtlety is not important for the discussion here.)
Including the $d=6$ operator, the charged current is given by 
\begin{equation}
J_\mu^{-} \equiv \bar e_{L \alpha} \gamma_\mu \left(U_{\rm eff}\right)_{\alpha i} \nu_i \, 
\end{equation}
with lepton mixing matrix $U_{\rm eff}$, whereas the neutral current is given by 
\begin{equation}
J_\mu = \frac{1}{2} \bar \nu_i \gamma_\mu \left( U_{\rm eff}^\dagger \right)_{i \alpha} \left(U_{\rm eff}\right)_{\alpha j} \nu_j \, ,
\end{equation}
which no longer equals $\frac{1}{2} \bar \nu_i \gamma_\mu \nu_i$, since $U_{\rm eff}$ is not unitary.

It turns out that leptogenesis already provides a very interesting and non-trivial constraint on the low-energy parameters $m_{\alpha \beta}$ and $\lambda_{\alpha \beta}$.  For example, one can show that leptogenesis vanishes if the $\lambda$ matrix eigenvalues are degenerate.\cite{bgj}  The leptogenesis constraint also requires nondegenerate $M_i$ and nondegenerate $m_i$ when the heavy neutrinos are strongly hierarchical.

The effects of the $d=6$ operator at low energy are presumably very tiny.  In the seesaw theory, the natural size of $\lambda$ is
of order $(v^2/2 M^2)$.  For the smallest seesaw scale which seems to be compatible with leptogenesis,\cite{DI}
$M \sim 10^8$~GeV, $\lambda \sim O(10^{-12})$, but it can be orders of magnitude smaller for larger $M$.  (For example, $M \sim 10^{14}$~GeV implies $\lambda \sim 10^{-24}$.) 
Probing $\lambda \sim 10^{-5}$ is achievable in next generation experiments.\cite{giudice}
Hence, the experimental situation is challenging in the extreme.  Nevertheless, bounding and eventually measuring the $d=6$ operator seems to be the only way to directly test the leading theory of leptogenesis.

The $d=6$ operator gives rise to some very interesting physical effects.
These include $CP$-violating asymmetries in neutrino-neutrino oscillations
as well as other effects which depend on the phases of the lepton sector.\cite{bgj}${}^,$\cite{gkm}
Even though experiments in the coming decade will yield in all probability null results, they are nonetheless important to pursue and putting experimental
bounds on the $d=6$ coefficients are of value.  For example, by bounding the $d=6$ coefficients, we can rule out (or in!)
more general theories with a scale of new physics of order the TeV scale, or electroweak scale.  Experiments which do this need to be part of the future experimental program in high-energy particle physics.

I think it is appropriate to recall the famous remark of Pauli when he first proposed the existence of the neutrino in order to preserve energy and momentum conservation in $\beta$ decay.  He said,
 ``I have done a terrible thing.  I have postulated a particle that cannot be detected."  While it is true
 that the neutrino could not be detected when it first was introduced by Pauli, direct experimental evidence
 for neutrinos was obtained 26 years later.  (Interestingly, Pauli was still alive!)
 This  experimental information was crucial for the development of the SM of particle physics.  
I think we are facing an analogous situation with the seesaw theory.  Experiment is very far from the sensitivity required to test leptogenesis.  Nevertheless, I believe that someday the experiments will be possible, and that coming generations (if not this generation) will unravel the mystery.

\section*{Acknowledgments}
I thank the organizers for the invitation to Coral Gables 2003.  This work was supported in part by
the Department of Energy under grant DE-FG03-97ER40546.

%
%
%
%

\end{document}